\documentclass[aps, prd, superscriptaddress, nofootinbib, twocolumn, preprintnumbers]{revtex4-2}
\usepackage[T1]{fontenc}

\usepackage{bm,amssymb,slashed,graphicx,multirow,soul,mathtools,xspace}
\usepackage{array}[=2016-10-06]
\usepackage{enumitem}
\usepackage{makecell}
\usepackage{placeins}
\usepackage{xcolor}

\definecolor{nicered}{rgb}{0.7,0.1,0.1}
\definecolor{nicegreen}{rgb}{0.1,0.5,0.1}
\definecolor{niceblue}{rgb}{0.1,0.1,0.5}

\usepackage[bookmarks=false]{hyperref}
\hypersetup{colorlinks,citecolor= nicegreen,linkcolor= niceblue,urlcolor= niceblue}

\newcommand{\g}{\gamma}

\makeatletter
\g@addto@macro\bfseries{\boldmath}
\makeatother

\allowdisplaybreaks
\begin{document}

\title{Precise $K^*(892) \to K\pi$ branching fractions}

\author{Vladimir V. Gligorov}
\affiliation{LPNHE, Sorbonne Universit{\'e}, Paris Diderot Sorbonne Paris Cit{\'e}, CNRS/IN2P3, Paris, France}
\author{Dean J. Robinson}
\affiliation{Ernest Orlando Lawrence Berkeley National Laboratory, 
University of California, Berkeley, CA 94720, USA}
\affiliation{Berkeley Center for Theoretical Physics, 
Department of Physics,
University of California, Berkeley, CA 94720, USA}

\begin{abstract}
Although discovered more than sixty years ago, direct measurement of the $K^*(892) \to K\pi$ branching fractions 
is a formidable challenge that has not been attempted.
Typically they are assumed to obey the isospin limit in hundreds of particle data measurements. 
We show that an abundance of recent amplitude analyses and other data, however, enables recovery of the ratios 
$\mathcal{B}(K^{*+} \to K^+ \pi^0)/\mathcal{B}(K^{*+} \to K_S^0 \pi^+)$ and 
$4\mathcal{B}(K^{*0} \to K_S^0 \pi^0)/\mathcal{B}(K^{*0} \to K^+ \pi^-)$ at $\sim5\%$ precision.
\end{abstract}

\maketitle

\textbf{Introduction.} In the isospin limit, the three branching ratios $\mathcal{B}(K^{*+} \to K^+ \pi^0) = 
\mathcal{B}(K^{*+} \to K^0_{S} \pi^+) = \mathcal{B}(K^{*+} \to K^0_{L} \pi^+) \simeq 1/3$
up to subpercent contributions from $K\g$ radiative decays (we write $K^*(892) = K^*$ hereafter).
Similarly $\mathcal{B}(K^{*0} \to K^+ \pi^-) = 4\mathcal{B}(K^{*0} \to K_{S}^0\pi^0)  = 4\mathcal{B}(K^{*0} \to K_{L}^0\pi^0) \simeq 2/3$.
Isospin-breaking effects are naively expected to arise at most at the few percent level,
per the characteristic breaking induced by light quark masses or electromagnetism.
While not a relevant source of phenomenological uncertainty today, these types of effects may 
become important at the ultimate sensitivities to quark-mixing processes targeted by the second 
upgrade of the LHCb detector, anticipated measurements by the BESIII and/or Belle~II experiments, 
and any upgrades thereof, and the planned FCC-$ee$ or CEPC $Z$-pole flavour programmes. It is 
therefore interesting to understand to what extent they can be directly constrained by experimental 
measurements.

Direct measurement of these $K^*$ decay modes would require a means to isolate 
the broad $P$-wave contribution from continuum effects, to reconstruct the $K_S\pi$ or $K^+\pi$ final states containing a $\pi^0$, 
and to accurately normalize the yield in order to extract a branching fraction.
The latter problem may be avoided by instead measuring ratios of branching fractions 
for ``$K^*$~cascade pairs'', 
i.e. pairs of  processes of the form $X \to Y (K^* \to K_S \pi)$ versus $X \to Y (K^* \to K^+ \pi)$,
though the problem of reliably estimating the different experimental efficiencies remains.
Thus such an approach has been historically highly nontrivial to achieve, 
and in the absence of such data over the intervening sixty four years since the $K^*(892)$ discovery~\cite{Alston:1961nx},
it has become the custom of experimental collaborations to use the isospin limit values.

The last twenty years, however, 
have seen a sizeable number of analyses involving the required $K^*$~cascade pairs, 
including a variety of amplitude analyses and other measurements~\cite{ParticleDataGroup:2024cfk}. 
While these analyses did not aim to measure $K^*$ branching fractions per se, we show that their results can be used for this purpose,
and that there are now enough such measurements that a combination can approach several percent precision.
Sensitivity to the isospin-breaking corrections in $K^* \to K \pi$ may be possible in the near future,
especially if experiments report individual product branching fractions involving the various $K^*$ modes 
they reconstruct. This would in turn allow the use of isospin limit values to be reconsidered.  

\begin{table}[t]
\newcolumntype{R}{ >{\raggedright\arraybackslash $} m{2.2cm} <{$}}
\newcolumntype{C}{ >{\raggedright\arraybackslash $} m{2.4cm} <{$}}
\newcolumntype{D}{ >{\centering\arraybackslash $} m{1.2cm} <{$}}
\newcolumntype{E}{ >{\centering\arraybackslash } m{1.5cm} <{}}
\renewcommand{\arraystretch}{1.25}
\resizebox{0.98\linewidth}{!}{\begin{tabular}{RCCDE}
\hline\hline
	\text{Process} & \mathcal{B} \!\times\! \mathcal{B}[K^{\pm} \pi^0] & \mathcal{B} \!\times\! \mathcal{B}[K^0_S \pi^{\pm}] & r_+ & Refs \\ 
\hline
	D^+ \to K^0_S K^{*+}	& 2.89(30) \times 10^{-3}	& {\!}^{\dagger}2.91(11) \times 10^{-3}	& 0.99(11)	& \cite{BESIII:2021dmo, BESIII:2024ncc, BESIII:2018pku} \\ 
	D^0 \to K^{*-} e^+ \nu	& 7.03(84) \times 10^{-3}	& {\!}^{\dagger}6.82(15) \times 10^{-3}	& 1.03(12)	& \cite{CLEO:2005cuk, BESIII:2024xjf} \\ 
	D^0 \to K^{*-} \mu^+ \nu	& {\!}^{\dagger}6.91(17) \times 10^{-3}	& 6.30(80) \times 10^{-3}	& 1.10(14)	& \cite{FOCUS:2004zbs, BESIII:2024qnx} \\ 
	D^0 \to K^{*-} \pi^+	& 2.31(30) \times 10^{-2}	& 1.64(15) \times 10^{-2}	& 1.41(22)	& \cite{CLEO:2000fvk, CLEO:2002uvu, BaBar:2008inr, ParticleDataGroup:2024cfk} \\ 
	D^0 \to K^{*+} K^-	& 1.52(8) \times 10^{-3}	& 1.89(30) \times 10^{-3}	& 0.80(13)	& \cite{BaBar:2007soq, LHCb:2015lnk, ParticleDataGroup:2024cfk} \\ 
	D^0 \to K^{*-} K^+	& 5.4(4) \times 10^{-4}	& 6.2(1.1) \times 10^{-4}	& 0.87(17)	& \cite{BaBar:2007soq, LHCb:2015lnk, ParticleDataGroup:2024cfk} \\ 
	D_s^+ \to K^{*+} K^0_S	& 2.04(33) \times 10^{-3}	& {\!}^{*}3.19(29) \times 10^{-3}	& 0.64(12)	& \cite{BESIII:2021anf, BESIII:2022npc, BESIII:2024oth} \\ 
	B^+ \to K^{*+} \pi^0	& 2.70(64) \times 10^{-6}	& 2.13(34) \times 10^{-6}	& 1.27(36)	& \cite{BaBar:2011cmh, BaBar:2015pwa} \\ 
	B^+ \to J\!/\!\psi K^{*+}	& 4.43(57) \times 10^{-4}	& 4.10(62) \times 10^{-4}	& 1.08(22)	& \cite{Belle:2002otd} \\ 
	B^0 \to K^{*+} \pi^-	& 2.67(45) \times 10^{-6}	& 2.46(16) \times 10^{-6}	& 1.09(20)	& \cite{Belle:2006ljg, BaBar:2009jov, BaBar:2011vfx, LHCb:2017pkd, ParticleDataGroup:2024cfk} \\ 
\hline\hline
	\text{Process} & \mathcal{B} \!\times\! \mathcal{B}[K^0_S \pi^0] & \mathcal{B} \!\times\! \mathcal{B}[K^{\pm} \pi^{\mp}] & r_0 & Refs \\ 
\hline
	D^+ \to \bar{K}^{*0} e^+ \nu	& {\!}^{\dagger}8.27(22) \times 10^{-3}	& 3.77(17) \times 10^{-2}	& 0.88(5)	& \cite{BaBar:2010vmf, BESIII:2015hty, BESIII:2024awg, ParticleDataGroup:2024cfk} \\ 
	D^+ \to \bar{K}^{*0} \pi^+	& 2.64(32) \times 10^{-3}	& 1.04(12) \times 10^{-2}	& 1.02(17)	& \cite{CLEO:2008jus, FOCUS:2009bwp, BESIII:2014oag, ParticleDataGroup:2024cfk} \\ 
	D^+ \to \bar{K}^{*0} \rho^+	& 9.7(9) \times 10^{-3}	& {\!}^{\dagger}4.15(18) \times 10^{-2}	& 0.93(10)	& \cite{BESIII:2023qgj, BESIII:2024waz, ParticleDataGroup:2024cfk} \\ 
	D^+ \to \bar{K}^{*0} K^+	& 5.2(1.4) \times 10^{-4}	& 2.49(10) \times 10^{-3}	& 0.84(23)	& \cite{CLEO:2008msk, BESIII:2021dmo, BESIII:2018pku, ParticleDataGroup:2024cfk} \\ 
	D^0 \to \bar{K}^{*0} \pi^0	& {\!}^{*}5.97(90) \times 10^{-3}	& 1.95(25) \times 10^{-2}	& 1.22(24)	& \cite{CLEO:2000fvk, CLEO:2011cnt, CLEO:2008kim} \\ 
	D_s^+ \to \bar{K}^{*0} K^+	& {\!}^{*}4.81(44) \times 10^{-3}	& {\!}^{*}2.62(5) \times 10^{-2}	& 0.73(7)	& \cite{BaBar:2010wqe, BESIII:2020ctr, BESIII:2022npc, BESIII:2024oth} \\ 
	D_s^+ \to K^{*0} \pi^+	& {\!}^{*}4.3(1.2) \times 10^{-4}	& {\!}^{*}1.68(26) \times 10^{-3}	& 1.02(33)	& \cite{FOCUS:2004muk, BESIII:2022vaf, BESIII:2021xox, BESIII:2024oth} \\ 
	B^+ \to K^{*0} \pi^+	& 1.68(33) \times 10^{-6}	& 6.69(54) \times 10^{-6}	& 1.01(21)	& \cite{Belle:2005rpz, BaBar:2008lpx, BaBar:2015pwa} \\ 
	B^0 \to J\!/\!\psi K^{*0}	& 2.00(42) \times 10^{-4}	& 8.67(87) \times 10^{-4}	& 0.92(21)	& \cite{Belle:2002otd} \\ 
\hline\hline
\end{tabular}}
\caption{$K^{*+,0}$ data and derived $r_{+,0}$ ratios.
For amplitude analyses, we cite the PDG~\cite{ParticleDataGroup:2024cfk} in the case that a normalization mode is obtained by a PDG fit.
References with weak pulls on combinations are not included.
Asymmetric uncertainties are averaged.
Results with updated normalizations are marked by an asterisk, and new results versus the PDG~\cite{ParticleDataGroup:2024cfk} by a dagger.
}
\label{tab:K*+0}
\vspace{-10pt}
\end{table}

\textbf{Data and results.} Table~\ref{tab:K*+0} lists those $K^*$ cascade pairs
that we are able to identify in the 2024 PDG listings~\cite{ParticleDataGroup:2024cfk},
as well as in several new results~\cite{BESIII:2024ncc, BESIII:2024xjf, BESIII:2024waz, BESIII:2024qnx, BESIII:2024awg},
from which one may obtain the ratios $r_+ \equiv \mathcal{B}(K^{*+} \to K^+ \pi^0)/\mathcal{B}(K^{*+} \to K_S^0 \pi^+)$ 
or $r_0 \equiv 4\mathcal{B}(K^{*0} \to K_S^0 \pi^0)/\mathcal{B}(K^{*0} \to K^+ \pi^-)$.
Other measurements may also constrain $r_{+,0}$, but their publications provide insufficient detail to unambiguously 
determine the contributions from the different  $K \pi$ final states.
In addition, we do not include measurements that reconstruct the $K^*$ solely via a mass window, 
such as in the $B \to K^*\gamma$ measurements~\cite{BaBar:2009byi, Belle:2017hum, Belle-II:2024kbc}.
When computing our averages we update several $D_s^+$ normalization modes affected by the recent Ref.~\cite{BESIII:2024oth},
which is not yet included in the PDG listings, and supersedes some previous BESIII results.

The cited measurements can be grouped into those which are limited by the available 
statistics~\cite{BESIII:2024ncc, CLEO:2005cuk, FOCUS:2004zbs, BESIII:2021anf, BaBar:2015pwa, Belle:2006ljg, CLEO:2011cnt, BESIII:2021xox}, 
those which are limited by systematic effects~\cite{BESIII:2021dmo, CLEO:2000fvk, CLEO:2002uvu, BaBar:2008inr, LHCb:2015lnk, CLEO:2008jus, FOCUS:2009bwp, BESIII:2023qgj, BESIII:2024waz, Belle:2005rpz, BaBar:2008lpx}, 
and those in which these two sources of uncertainty are of roughly equal size~\cite{BESIII:2018pku, BESIII:2024xjf, BESIII:2024qnx, BaBar:2007soq, BESIII:2022npc, BESIII:2024oth, BaBar:2011cmh, BESIII:2014oag, BaBar:2009jov, BaBar:2011vfx, LHCb:2017pkd, BaBar:2010vmf, BESIII:2015hty, BESIII:2024awg, CLEO:2008msk, CLEO:2008kim, BaBar:2010wqe, BESIII:2020ctr, FOCUS:2004muk, BESIII:2022vaf, Belle:2002otd}.
As the majority of the cited measurements fall into the latter two categories, combining them
requires assumptions about potential correlations in the methods used to evaluate systematic 
uncertainties. 

This is especially relevant for amplitude analyses
in which the chosen amplitude model and lineshape parameters are a notable source of 
systematic uncertainty. Unfortunately, none of the cited analyses report correlations between their
systematic uncertainties, and only one~\cite{LHCb:2015lnk} reports a full covariance matrix for the
measured physical observables including both statistical and systematic effects. Around a third
do not report the relative contribution of individual systematics
to the total uncertainty, and none report the sign of the shift in the measured physical
observables associated with any given systematic uncertainty, even though in several cases the 
total reported systematic uncertainty is highly asymmetric. 

Particularly problematic is the treatment of meson radii, which are parameters of the resonant
lineshapes that typically have to be fixed in amplitude analysis fits in order to ensure a stable convergence.
These radii are frequently among the leading sources of uncertainty for the fit fractions in our
cited amplitude analyses, yet we find in e.g. Refs~\cite{BESIII:2021dmo,BESIII:2024ncc,BESIII:2022vaf,BESIII:2021xox, LHCb:2017pkd, BaBar:2010vmf, BESIII:2023qgj, BESIII:2024qnx,LHCb:2015lnk} differences in both the chosen nominal
values for these parameters as well as in the range of the parameters considered as an appropriate 
systematic variation. Several of the analyses make fully disjoint choices, such that the nominal
value of one analysis is not even within the range of systematic uncertainty considered by another.
Similar comments could be made about the choice of fit components in the different amplitude analyses
which enter our combination,
and furthermore, 
systematic effects arising from track-finding and particle identification and reconstruction algorithms may be correlated across measurements.
It is entirely possible that a coherent analysis of these datasets 
would change both the average central values and their uncertainties, 
but there is no way to quantify this supposition given the information reported in these analyses. Given the available information, 
we combine measurements assuming that the reported systematic uncertainties are fully uncorrelated.

\def\resrp{0.95 \pm 0.06}
\def\resrz{0.87 \pm 0.03}
\def\resrpnopull{1.01(5)}
\def\resrznopull{0.97(7)}
\def\resrzSL{0.93(3)}
\def\resrzSLav{0.91(3)}
\def\Svalrp{1.3}
\def\Svalrz{1.0}
\def\Svalrpnopull{1}
\def\Svalrznopull{1}
\def\SvalrzSL{1}
\def\SvalrzSLav{1.1}
\def\Soprp{=}
\def\Soprz{=}
\def\Soprpnopull{<}
\def\Soprznopull{<}
\def\SoprzSL{<}
\def\SoprzSLav{=}

With these caveats in mind, we proceed to combine the $r_+$ and $r_0$ results similarly assuming that their uncertainties are fully uncorrelated. This yields 
\begin{equation}
	\label{eqn:rpz}
	r_+ = \resrp\,, \quad \text{and} \quad  r_0 = \resrz\,,
\end{equation} 
including a PDG-style uncertainty rescaling `$S$ factor' of $S \Soprp \Svalrp$ and $S\Soprz \Svalrz$, respectively.

Downward pulls arise in the averages for both $r_+$ and $r_0$ from $D_s^+$ data.
In the case of $r_0$, the amplitude analyses for $D_s^+ \to (\bar{K}^{*0} \to K^- \pi^+)K^+$~\cite{BaBar:2010wqe, BESIII:2020ctr}  
agree well with each other. 
While there is only one analysis for $D_s^+ \to (K^{*+} \to  K_S^0 \pi^+)K_S^0$~\cite{BESIII:2021anf},
it nevertheless seems plausible
that the $D_s^+ \to K_S^0 K^+ \pi^0$ amplitude analysis~\cite{BESIII:2022npc}, 
which affects both $r_+$ and $r_0$, 
may be the sole source of the observed tension from $D_s^+$ data in our averages.
Further, a moderate downward pull arises in $r_0$ from $D^+ \to \bar{K}^{*0} e^+ \nu$.
We note a significant tension in the PDG fit for the $D^+ \to K^- \pi^+ e^+ \nu$ normalization mode, 
measured by Refs.~\cite{BaBar:2010vmf, BESIII:2015hty}: 
using only the direct measurement~\cite{BESIII:2015hty} one finds for this mode $r_0 = \resrzSL$
and the weighted average becomes $r_0 = \resrzSLav$ with $S \SoprzSLav \SvalrzSLav$.
If we exclude the $D_s^+ \to \bar{K}^{*0} K^+$, $D_s^+ \to K^{*+} K_S^0$ and $D^+ \to \bar{K}^{*0} e^+ \nu$ ratios from the weighted averages,
the obtained results become $r_+ = \resrpnopull$ and $r_0 = \resrznopull$.

\textbf{Conclusions.} We have demonstrated that existing measurements of processes involving
the charged and neutral  $K^*(892)$ resonances allow their $K\pi$ branching fractions 
to be constrained at the level of a few percent. 
We anticipate that the much larger datasets to be collected by ongoing and planned flavour experiments
will allow percent-level determinations,
provided systematic effects can be controlled. 
We have highlighted the way in which correlated assumptions in the
different analyses contributing to the $r_{0,+}$ combinations may complicate their precise evaluation.
These difficulties can be overcome by performing coherent combined analyses of the various experimental
datasets which have been, or will be, collected. 
In the meantime, these concerns, combined with large pulls in the weighted averages,
warrant caution in the interpretation of any apparent sizeable tensions with the isospin limit.

The $K^*(892)$ branching fractions considered here may be 
just one example of the way in which flavour datasets can be leveraged to improve our knowledge of 
fundamental interactions, 
beyond the headline measurements that the experiments are ostensibly built to perform. 
This communication therefore motivates ongoing work to facilitate joint analyses of
these experimental datasets in order to unlock their full scientific potential.

\textbf{Acknowledgements.} 
We thank Fr\'ed\'eric Blanc, Tim Gershon, Zoltan Ligeti, Marco Pappagallo, Ezra Robinson, and Piotr Zyla for discussions and comments on this manuscript.
We thank the generations of experimentalists whose tireless efforts enabled this analysis.
DJR is supported by the Office of High Energy Physics of the U.S. Department of Energy under contract DE-AC02-05CH11231. 

\vfill

\end{document}